          [1999/12/02 1.01 (PWD)]
\documentclass[a4paper,twocolumn,]{esapub} % European paper
\input{epsf}
\usepackage{graphicx}
\usepackage{natbib}

\title{Innermost stable 
circular orbits around rotating compact quark stars and QPOs}
\author{D. Gondek-Rosi\'nska}
\author {T. Bulik}
\author {W. Klu\'zniak}
\author {J. L. Zdunik}
\affil{Copernicus Astronomical Center, Bartycka 18, Warsaw,Poland }
\author{E. Gourgoulhon}
\affil{DARC, UMR 8629 du CNRS, Observatoire de Paris, F-92195 Meudon  Cedex,
 France}

\begin{document}


\maketitle

\begin{abstract}
  It has been suggested that observations of quasi periodic
  oscillations (QPOs) in the X-ray fluxes from low mass X-ray binaries
  can be used to constrain the mass of the compact object and the
  equation of state of its matter.  A specific suggestion that the kHz QPO
  frequency saturates at the maximum orbital frequency has been
  widely considered.  We examine rotating compact stars described by a
  new model of strange quark matter (Dey et al.  1998). We calculate
  the maximum orbital frequencies for both normal and supramassive 
  strange stars described by the Dey et al.
  model, and present these frequencies
  for sequences of equilibrium models with constant baryon mass.
   The maximum orbital frequencies for these compact objects
  are always higher than the kHz QPO frequencies observed to date.

\end{abstract}

\section{Introduction}
The recently discovered kHz QPOs can be used as a probe of the inner
regions of accretion disks in compact stars and hence also of the
properties of the central object.  These oscillations have been used to
derive estimates of the mass of the neutron star in some sources
(e.g. Kaaret et al. 1997, Zhang et al. 1998, Klu\'zniak 1998) and of
strange stars given by the MIT bag model (Bulik et al. 1999,
Stergioulas et al. 2000, Zdunik et al 2000a,b).  All these estimates
assume that the maximum observed frequency is the orbital
frequency in the innermost marginally stable orbit,
as suggested by Klu\'zniak et al. (1990). 

Very recently, a new model of
strange quark matter (Dey et al.  1998) was used for calculating
frequencies of marginally stable orbits around static strange stars
and strange stars rotating with frequency 200 Hz and 580 Hz (Datta et al.
2000). The authors conclude that very high QPO frequencies in the
range of $1.9$-$3.1$\, kHz, if observed,
 would imply the existence of a compact
strange star in the X-ray binary, rather than a neutron star.
Two cases
of the Dey model have been used in these papers. 
 Both cases (differing in the value of a free parameter)
give a rather low value
for the maximum gravitational mass of static
configurations $M_{max}=1.32\,M_\odot$ (with the stellar radius $R=6.5$~km)
and $M_{max}=1.44\,M_\odot $ ($R=7.1$~km).

  In the present paper we
calculate the highest Keplerian frequencies 
of rotating quark stars, for two Dey models, for equilibrium sequences
with fixed baryon mass. We perform calculations
for all possible stellar rotation rates.
 We find the absolute upper limit on the orbital
frequency. We show that the Dey models considered do not allow
maximum frequencies which are lower than 1.5 kHz for
 stars with masses $M>1M_\odot$.

\section{Calculations} 

We have calculated the innermost stable circular orbit of the
uniformly rotating strange stars described by the Dey model
using the multi-domain spectral methods developed
by Bonazzola et al. (1998). This method has been used previously for
calculating rapidly rotating strange stars described by the MIT bag
model (Gourgoulhon et al. 1999) and for
finding basic properties of rapidly rotating strange stars
described by the Dey model (Gondek-Rosi\'nska et al. 2000). The
multi-domain technique allows one to address the density discontinuity
at the surface of self-bound stars. 

 The Dey et al. model (in contrast to MIT
bag model) describes quark confinement self-consistently. The
mean density of the stars considered here is about $ 10^{15}\,$g/cm$^3$.
The stars are very compact---the
gravitational redshifts $z$ for the maximum-mass static configurations
are much larger than those of strange stars within the MIT bag model
or of most models of neutron stars.  Thus, Dey stars can rotate
extremely fast.  We perform calculations
for equilibrium sequences with baryon masses from $0.5M\odot$
upwards. The maximal baryon
mases are  $1.64\ M_\odot$ and $1.85\ M_\odot$
for static Dey stars in the two cases
considered (Gondek-Rosi\'nska et al. 2000), but we include supramassive
stars with baryon masses
up to $2.2\ M_\odot$ and $2.5\ M_\odot$, respectively. The results
for the two models  are plotted
in Fig. 1 and Fig. 2, respectively.

\section{Orbital frequencies}

We find that: 

a) for stars with moderate and high baryon masses
(higher than $\sim 60\ \%$ of the maximum baryon mass of static
configurations) a gap always separates the innermost stable
circular (ISCO) and the stellar surface, for any
stellar rotation rate; 

b) for low mass-stars  with moderate rotation rates,
stable orbits extend down to the stellar surface
 (see, e.g.,  in Fig 1. the parts between the cusps of the sequences with
baryon masses  0.87 and $0.79M_\odot$);

c) in the Dey model, as in the MIT bag model,
for strange stars of any mass rotating at the equatorial
mass-shedding limit, a gap always separates the ISCO and the stellar
surface;

d) depending on the baryon mass of the star,
  the lowest ISCO frequency is attained either in the static configuration
 or in the configuration at the equatorial mass-shedding limit
 (see the figures);

e) the range of ISCO frequencies for Keplerian models is fairly narrow,
and the dependence of ISCO
frequency on the  rotational frequency is well aproximated by a linear
function $f_{\rm ISCO} \approx 0.8f$; 

f) the least upper bound on the orbital frequency of stars modeled
with the  Dey et al. (1998) equation of state
 is obtained for a non-rotating star (the one whose radius
is equal to the ISCO radius).

\section{Implications for QPO models}

If some of the kHz QPO sources are Dey strange stars, then the QPO frequency
has to be significantly below the maximum orbital frequency.
In fact, in the discussion of QPOs in black hole candidates,
the QPO frequency is taken to be the trapped mode frequency
(Nowak, Wagoner 1990, Nowak et al. 1997),
and it is seems worthwile to explore further the possibility that
the highest frequency QPOs may appear 
at suborbital frequencies also in other systems, such as neutron
star binaries and in AGNs.

\begin{figure*}
\centering
\leavevmode
%\epsfxsize=15cm
%\epsfysize=15cm
%\epsfbox{fig1.eps}
\includegraphics[width=0.8\linewidth]{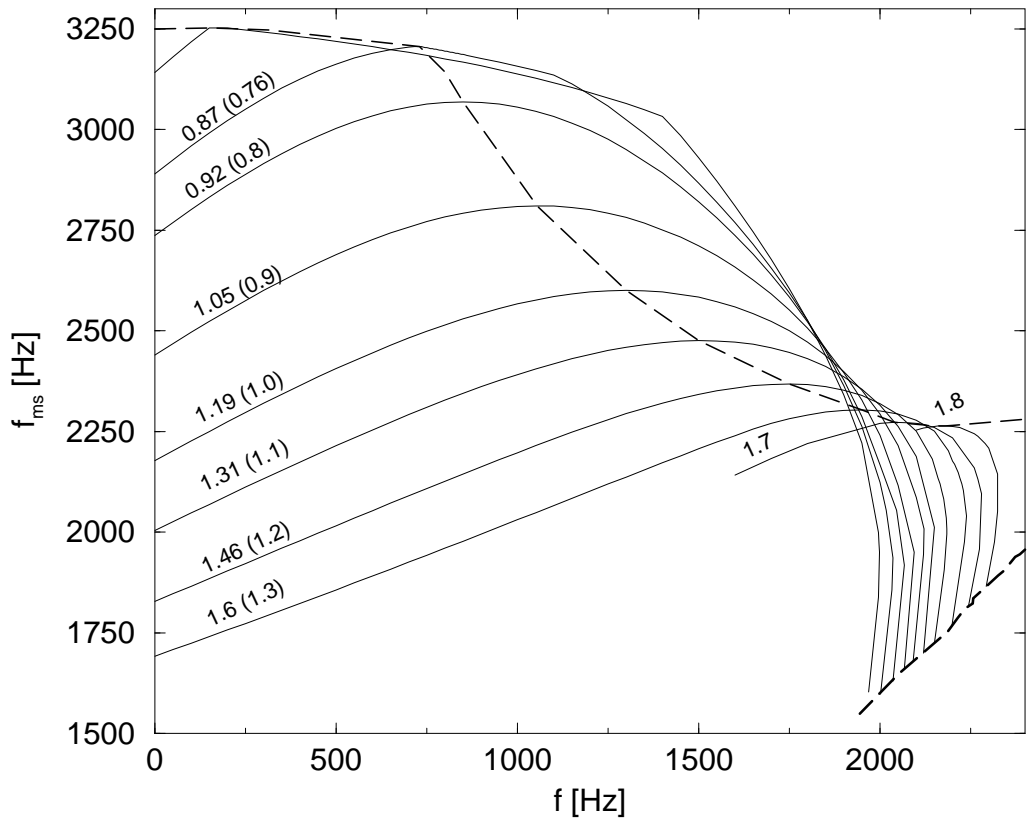}
\caption{Frequency of the innermost marginally stable orbit as a function 
  of stellar spin rate for one of the Dey models (in this model the maximum
 gravitational mass of a non-rotating, i.e., static star is  $1.32 M_\odot$). 
 The sequences shown with thin continuous lines are labelled with their 
  fixed baryon mass, in  solar masses
 (the number in parentheses is the
  gravitational mass of the static configuration).  The angular
  momentum increases along each curve, up to the Keplerian configuration,
  indicated with the thick dashed line. The thin dashed line connects
  the maxima of the curves.
   }
\label{fig1}
\end{figure*}
\begin{figure*}
\centering
\leavevmode
\includegraphics[width=0.8\linewidth]{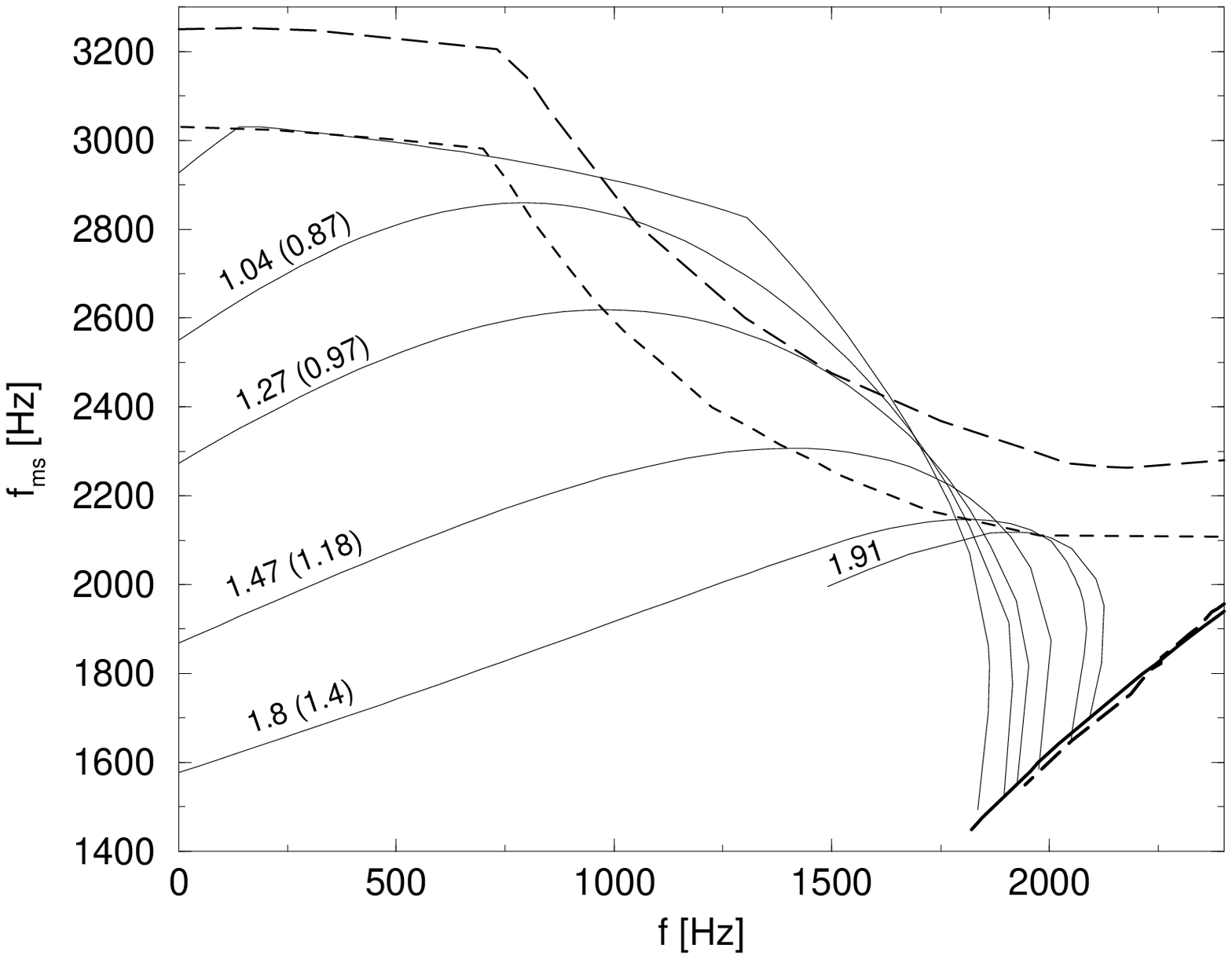}
\caption{Same as Fig. 1, for the other Dey model (here, the maximum
  gravitational mass of a static star is  $1.44 M_\odot$). 
   The mass-shed limit for this model is indicated with the
    thick solid line (and the corresponding limit for the model of Fig. 1
   is shown again, for comparison, as a thick long-dashed line).
The maxima of the curves are connected with the short-dashed line,
and the corresponding thin long-dashed 
line from Fig. 1 is also reproduced for comparison.
     }
\label{fig2}
\end{figure*}
\section*{ACKNOWLEDGMENTS}
This work has been funded by the KBN grants 2P03D01816, 2P03D00418, 
2P03D01413. The numerical calculations have been performed on
computers purchased thanks to a special grant from the SPM and SDU
departments of CNRS.
\section*{REFERENCES}

Bombaci, I., Datta, B., 2000, ApJ, 530, L69

Bonazzola, S., Gourgoulhon, E., Marck, J. A. 1998, Phys. Rev. D, {58}, 104020

Bulik, T., Gondek-Rosi{\'n}ska, D., and Klu\'zniak, W., 1999a, A\&A, 344, L71

Datta, B., {Thampan}, A.~V., and {Bombaci}, I., 2000,
 {355}, L19

Dey, M., {Bombaci}, I., {Dey}, J., {Ray}, S., and {Samanta}, B.~C., 1998,
Physics Letters B, {438}, 123

Gondek-Rosi{\'n}ska, D., Bulik T., Zdunik J. L., Gourgoulhon E., Ray S.,
 Dey J., Dey M., 2000, A\&A, 363, 1005

Gourgoulhon, E., {Haensel}, P., {Livine}, R., {Paluch}, E., {Bonazzola}, S.,
  and {Marck}, J.~A., 1999, A\&A, {349}, 851

Kaaret, P., Ford, E. C., Chen, K. 1997, ApJ, 480, 127

Klu\'zniak, W., Michelson, P., Wagoner, R. V., 1990, ApJ, 358, 538

Klu\'zniak, W., 1998, ApJ, 509, L37

Nowak, M. A., Wagoner, R. V., 1991, ApJ, 378, 656

Nowak, M.A, Wagoner, R. V. , Begelman, M.C., Lehr, D.E.,
  1997, ApJ, 477, L91

Stergioulas, N., Klu{\'z}niak, W., and Bulik, T., 1999, A\&A, 352, L116

Zhang, W., Strohmayer, T. E., Swank, J. H. 1997, ApJ, 482, L167

Zdunik, J.\ L., Bulik, T., Klu{\'z}niak, W., Haensel, P., 
Gondek-Rosi{\'n}ska, D.\ 2000a, A\&A, 359, 143 

Zdunik, J. L., {Haensel}, P., {Gondek-Rosi{\'n}ska}, D., {Gourgoulhon}, E.,
 2000b, {A\&A} {356}, 612 

\end{document}